\documentclass[a4paper,11pt]{article}
\usepackage[margin=2.5cm]{geometry}
\usepackage{setspace,graphicx,url,hyperref,enumitem,multirow}
\usepackage{amsmath,amssymb,amsfonts,amsthm}
\usepackage{color,colortbl}
\definecolor{Gray}{gray}{0.9}

\newcommand{\R}{\ensuremath{\mathbb{R}}}

\begin{document}
\onehalfspacing
\title{Using machine learning for quantum annealing accuracy prediction}
\author{Aaron Barbosa, Elijah Pelofske, Georg Hahn, and Hristo N.\ Djidjev}
\date{Los Alamos National Laboratory}
\maketitle

\begin{abstract}
	Quantum annealers, such as the device built by D-Wave Systems, Inc., offer a way to compute solutions of NP-hard problems that can be expressed in Ising or QUBO (quadratic unconstrained binary optimization) form. Although such solutions are typically of very high quality, problem instances are usually not solved to optimality due to imperfections of the current generations quantum annealers. In this contribution, we aim to understand some of the factors contributing to the hardness of a problem instance, and to use machine learning models to predict the accuracy of the D-Wave 2000Q annealer for solving specific problems. We focus on the Maximum Clique problem, a classic NP-hard problem with important applications in network analysis, bioinformatics, and computational chemistry. By training a machine learning classification model on basic problem characteristics such as the number of edges in the graph, or annealing parameters such as D-Wave's chain strength, we are able to rank certain features in the order of their contribution to the solution hardness, and present a simple decision tree which allows to predict whether a problem will be solvable to optimality with the D-Wave 2000Q. We extend these results by training a machine learning regression model that predicts the clique size found by D-Wave.
\end{abstract}

\section{Introduction}
\label{sec:introduction}
Quantum annealing is an emerging technology with the potential to provide high quality solutions to NP-hard problems. In this work, we focus on the devices built by D-Wave Systems, Inc., specifically the D-Wave 2000Q annealer, designed to minimize functions of the following form,
\begin{align}
    H(x_1,\ldots,x_n) = \sum_{i=1}^N h_i x_i + \sum_{i<j} J_{ij} x_i x_j,
    \label{eq:hamiltonian}
\end{align}
where $x_i$ are unknown binary variables. The linear weights $h_i \in \R$ and the quadratic coupler weights $J_{ij} \in \R$ are chosen by the user and define the problem under investigation. If $x_i \in \{0,1\}$, eq.~\eqref{eq:hamiltonian} is called a \textit{quadratic unconstrained binary optimization (QUBO)} problem, and if $x_i \in \{-1,+1\}$ it is called an \textit{Ising model}. Both QUBO and Ising model formulations are equivalent \cite{Chapuis2019}. Many important NP-hard problems can be expressed as the minimization of a function of the form of eq.~\eqref{eq:hamiltonian}, see \cite{Lucas2014}. In the remainder of the article, we focus on the QUBO formulation.

Although solutions returned by the D-Wave 2000Q annealer for minimization problems of the type of eq.~\eqref{eq:hamiltonian} are typically of very high quality, they are not guaranteed to be optimal. In fact, there is no guarantee that the solutions returned by a D-Wave device keep any level of accuracy.

In this article, we aim to identify those types of QUBOs that are solvable on D-Wave 2000Q with given hardware parameters, without actually using the annealer. This could help decide on what resources (annealing time, number of reads) should be allocated depending on the problem hardness. We focus on the Maximum Clique (MC) problem, and explore two different types of classifications. First, we train a binary machine learning classification model on basic problem characteristics (such as density of the input graph) to discriminate between solvable and unsolvable instances of MC. The resulting model also allows us to rank the features used for training by importance. Classification is performed using the open source machine learning package \textit{scikit-learn} in Python \cite{scikit-learn}. Second, we show that the decision problem of whether an MC instance will be solved optimally by D-Wave can be predicted with high accuracy by a simple decision tree on the same basic problem characteristics. Third, we train a machine learning regression model to predict the clique size returned by the annealer.

\begin{figure}[t]
\centering
    \includegraphics[width=0.5\columnwidth]{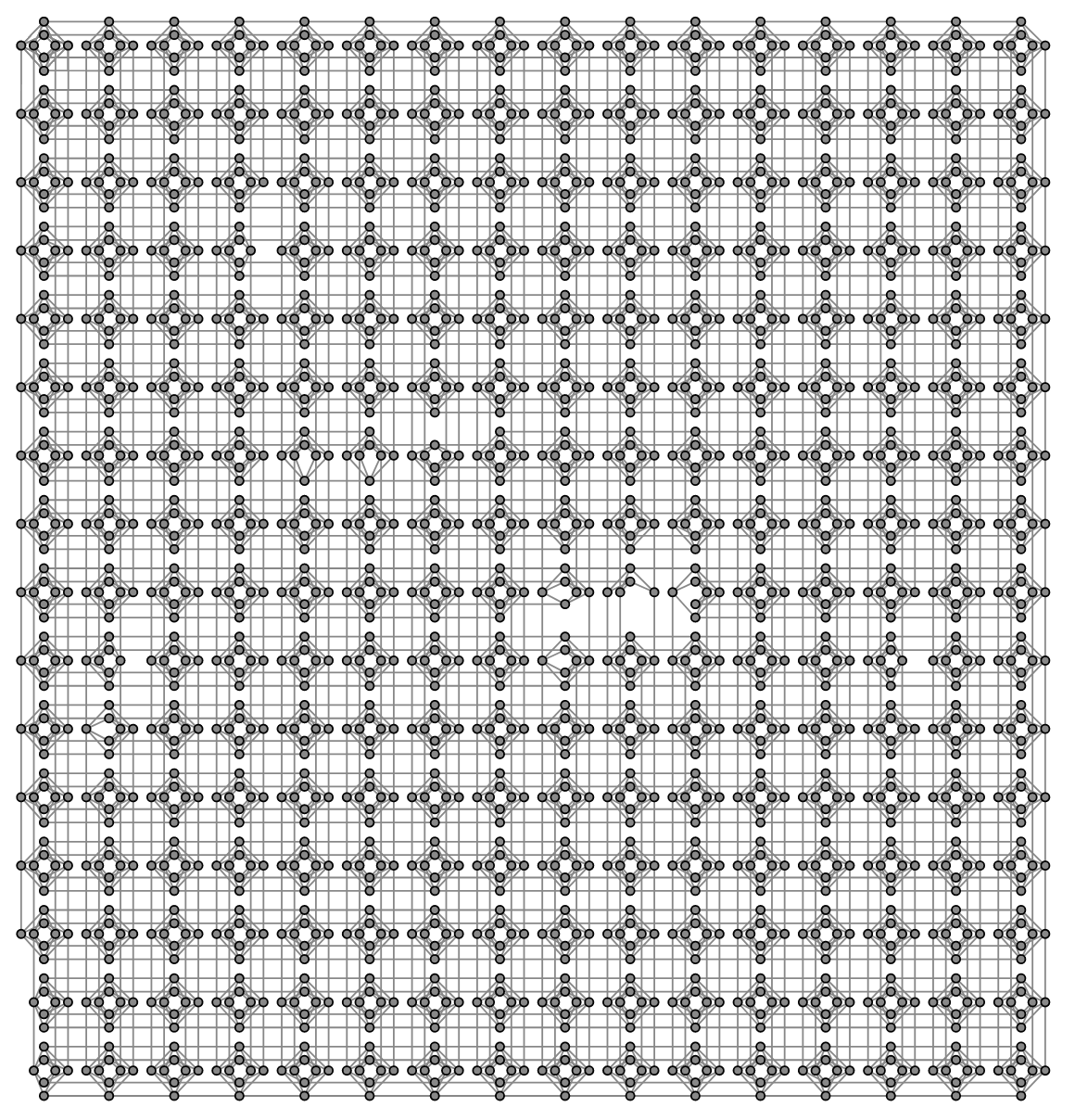}
    \caption{The hardware graph $H$ of  D-Wave 2000Q is a $12 \times 12$ array of unit cells, wherein each unit cell is a $4 \times 4$ bipartite graph. The depicted graph is the hardware connectivity graph of the D-Wave 2000Q machine at Los Alamos National Laboratory, showing a few of its qubits disabled due to hardware errors. The particular connectivity structure of the hardware graph is referred to as a \textit{Chimera graph}.\label{fig:chimera}}
\end{figure}

Let $G=(V,E)$ be a graph (referred to as \textit{original graph}) consisting of a vertex set $V \subseteq \{1,\ldots,n\}$ and edge set $E \subseteq V \times V$. A \textit{clique} in $G$ is any subgraph $C$ of $G$ that is \textit{complete}, i.e., there is an edge between each pair of vertices of $C$. The MC problem asks us to find a clique of maximum size, called a \textit{maximum clique}, which is an NP-hard problem with many applications. To solve an instance of MC on the D-Wave annealer, we proceed as follows:
\begin{enumerate}
    \item The task of finding a maximum clique has to be mapped to a formulation of the type of eq.~\eqref{eq:hamiltonian}. As shown in \cite{Chapuis2019,qtop}, the QUBO function
    \begin{equation}
        H = -A\sum_{i \in V} x_i + B\sum_{(i,j ) \in \overline{E}} x_i x_j,
        \label{eq:MC}
    \end{equation}
    $x_i\in\{0,1\}$ for all $i \in V$, achieves a minimum if and only if the vertices $i \in V$ for which $x_i=1$ form a maximum clique, where the two constants can be chosen as $A=1$, $B=2$ \cite{qtop}. It is noteworthy that any function of the type of eq.~\eqref{eq:hamiltonian} can be represented as a graph $P$ itself, which is used to embed the QUBO problem into the quantum processor. In this representation, each of the $n$ variables $x_i$ becomes a vertex with vertex weight $h_{i}$, and each edge between vertices $i$ and $j$ is assigned an edge weight $J_{ij}$. We call $P$ the \textit{unembedded graph}.
    \item To minimize eq.~\eqref{eq:hamiltonian} on D-Wave, its corresponding unembedded graph has to be mapped onto the physical hardware qubits on the chip of the D-Wave 2000Q. Since the connectivity of the hardware qubits, called the Chimera graph (see Figure~\ref{fig:chimera}), is limited, a \textit{minor embedding} of $P$ onto the Chimera graph has to be computed. In a minor embedding, each vertex of $P$ (referred to as a \textit{logical qubit}) is mapped onto a connected set of hardware qubits, called a \textit{chain}. As such, a coupler weight, called a \textit{chain strength}, has to be specified for each pair of connected chain qubits. If chosen appropriately, the chain strength ensures that hardware qubits representing the same logical qubit take the same value after annealing is complete. The minor embedding of $P$ onto the Chimera graph is a subgraph of the Chimera graph and thus a new graph $P'$. We call $P'$ the \textit{embedded graph}.
    \item After annealing, it is not guaranteed that chained qubits on the Chimera graph take the same value, even though they represent the same logical qubit. This phenomenon is called a \textit{broken chain}. To obtain an interpretable solution of logical qubits, definite values have to be assigned to each logical qubit occurring in eq.~\eqref{eq:hamiltonian}. Although there is no unique way to accomplish this task, D-Wave offers several default methods to \textit{unembed} chains.
\end{enumerate}

Applying machine learning to predict features of a quantum device is a timely area of research. Existing work mostly focuses on gate quantum computing. For instance, machine learning is employed to help tune the behavior of quantum systems, such as computing the overlap of two quantum states \cite{Cincio2018}, quantum circuit optimization \cite{Verdon2019, Fosel2021}, or readout of states with higher assignment fidelities under noise conditions \cite{Magesan2015}. Other works address parameter tuning of variational quantum algorithms \cite{Rivera2021} or the QAOA (quantum approximate optimization algorithm) algorithm of \cite{farhi2017quantum}, see \cite{Wauters2020}. The closest to our approach is \cite{Moussa2020}, wherein the authors use machine learning techniques to identify graph problems that are easy to solve using QAOA. However, they use the quantum gate model and their specific objective is a bit different---to decide whether QAOA or the classical Goemans-Williamson algorithm will perform better on instances of the Maximum Cut problem. In contrast, we compare the quality of the solution found with the quantum algorithm against an optimal solution, and our optimization problem is the Maximum Clique problem.

Somewhat related, in \cite{spin-reversal, Barbosa} the authors use a differential evolution optimization to tune three parameters of D-Wave 2000Q, namely spin reversal, anneal offsets, and chain weights. However, machine learning methods are not used, and the objective is to optimize parameters, rather than predict performance.

This article is structured as follows. Section~\ref{sec:methods} introduces the machine learning methodology we employ to predict the behavior of D-Wave 2000Q. Experimental results are given in Section~\ref{sec:experiments}, where we consider the classification problem of predicting whether D-Wave 2000Q can solve a particular MC instance, the computation of a simple decision tree allowing us to classify problem instances manually, and the prediction of the clique size returned by the annealer using a machine learning regression model. The article concludes with a discussion in Section~\ref{sec:discussion}.

\section{Methods}
\label{sec:methods}
We aim to predict both if an instance of MC is solvable by D-Wave 2000Q to optimality, as well as the size of the clique which will be found by the annealer.

As test and validation datasets, we first generate a set of Erd\H{o}s-R\'enyi random graphs with a number of vertices in $\{20,\ldots,65\}$, and a density chosen uniformly at random in $[0.01,0.99]$. The upper bound of $65$ vertices is determined by the largest size of a complete graph embeddable on D-Wave's Chimera architecture. We specifically do not use any graphs with zero degree nodes. Next, we compute an embedding of a complete 65 node graph onto the Chimera architecture with the help of the \textit{minorminer} module of the D-Wave Ocean API \cite{minorminer}. In order to be able to compare and interpret results across different graphs, we keep the embedding fixed in all experiments. The chain strength for the embedding is always computed using D-Wave's \textit{uniform torque compensation} feature \cite{dwave_torque}. The best D-Wave result (i.e., the largest clique found) is then determined among $1000$ anneals. We use the majority vote technique to unembed all broken chains (in particular, using an unbiased coin flip if a chain is balanced). The split ratio between test and validation datasets is always chosen as $90\%$ (validation) and $10\%$ (testing).

\subsection{Classification}
Our task is to relate graph features to a given binary indicator from D-Wave expressing if an instance could be solved by the annealer to optimality. Several avenues exist to accomplish this task, for instance via (penalized) linear regression or logistic regression, or via more sophisticated machine learning models. The following highlights some of the challenges.

First, in order to have a well-defined decision problem for finding maximum cliques, we require a ground truth determined with a classical method. We define an instance of MC to be solvable by the D-Wave 2000Q if the clique determined by D-Wave has size equal to the maximum clique size found with the function \textit{graph\_clique\_number} of the \textit{NetworkX} package in Python \cite{networkx, graphcliquenumber}, which is an exact solver.

Second, a couple of choices have to be made, both regarding the machine learning model for regression, as well as the set of graph features selected for prediction. We decided to use a decision tree classifier for two main reasons: The classifier achieved good performance in the classification task we consider and, most importantly, it allows us to obtain an interpretable output in the form of a decision tree. The decision tree highlights in what order/importance the features contribute to solvability on the D-Wave 2000Q, and by tuning the decision tree for simplicitly, it allows one to (manually) determine with high probability in advance if an instance is likely solvable.

Third, an assortment of graph-related features has to be selected to serve as inputs to the machine learning model. Those features are selected to cover a wide variety of potential metrics impacting solvability, however the list below is neither exhaustive nor rigorously proven. The following features went into the machine learning model threefold (unless noted otherwise), precisely for the original input graph $G$, its unembedded graph $P$ of the MC problem, and its embedded graph $P'$ on the Chimera architecture (see Section~\ref{sec:introduction}):
\begin{enumerate}
    \item the graph density (\textit{Graph\_Density});
    \item the minimal, maximal, and average degree of any vertex (\textit{Graph\_Min\_Degree},\linebreak \textit{Graph\_Max\_Degree}, and \textit{Graph\_Mean\_Degree});
    \item the number of trianges in the input and unembedded graphs (\textit{Graph\_Num\_Triangles});
    \item the number of nodes and edges (\textit{Graph\_Num\_Nodes} and \textit{Graph\_Num\_Edges});
    \item the $N \in \{1,\ldots,5\}$ largest eigenvalues of the adjacency matrix (\textit{Graph\_Largest\_Eigenvalue} and \textit{Graph\_Nth\_Largest\_Eigenvalue}), and the spectral gap (the difference between the moduli of the two largest eigenvalues) of the adjacency matrix (\textit{Graph\_Spectral\_Gap}). For brevity of notation, we refer with "eigenvalue of a graph" to the eigenvalue of its adjacency matrix.
\end{enumerate}

Additionally, for the embedded graph $P'$ on the Chimera architecture, we included the following features into the model:
\begin{enumerate}
    \item the minimal, maximal, and average chain length occurring in the embedding\linebreak (\textit{Min\_Chain\_Length}, \textit{Max\_Chain\_Length}, and \textit{Avg\_Chain\_Length});
    \item the chain strength computed with D-Wave's \textit{uniform torque compensation} feature\linebreak (\textit{Chain\_Strength}), see \cite{dwave_torque}, using either a fixed UTC prefactor (Section~\ref{sec:fixedsetting}) or a randomly selected one in $[0.5,3]$ (Section~\ref{sec:randomsetting});
    \item the annealing time (in microseconds), which was selected uniformly at random in $[1,2000]$ (\textit{Annealing\_Time}).
\end{enumerate}
In total, there were $46$ features that were included as inputs to the machine learning model.

Using those features of the training MC instances and their maximum clique result computed by the D-Wave 2000Q annealer, we train a machine learning classifier implemented in the \textit{sklearn.tree.DecisionTreeClassifier()} class provided in \textit{scikit-learn} \cite{scikit-learn}. After performing a grid search across the following parameters, we selected \textit{max\_depth=5}, \textit{random\_state=0}, and \textit{min\_impurity\_decrease=0.005}. All other parameters were kept at their default values. To weigh solvable MC instances by D-Wave more heavily than unsolvable ones, the option \textit{class\_weight='balanced'} was employed. The option of balanced class weightings was chosen as only about $p \approx 0.11-0.13$ of the test problems are solvable. Therefore, a model that classifies all instances as unsolvable will still be $87-89\%$ accurate. By setting \textit{class\_weight='balanced'}, we assign a weight to the solvable problems that is inversely proportional to how often they appear in the dataset, thus penalizing incorrectly classified solvable problems by a factor of about $1/p$.

During training, we noticed that the decision tree machine learning model appears to have trouble with problems that return a clique size of zero. This was due to a suboptimal QUBO formulation we used for MC, and the problem disappeared when using the formulation of \cite{Chapuis2019, qtop}. However, similar issues might occur with other problems, highlighting that the present approach might be most suitable for problem classes that can be solved sufficiently well on D-Wave. Additionally, it is challenging to tune the decision tree in such a way as to obtain good performance and interpretable results simulatenously. Applying the decision tree classifier using default parameters usually results in very large trees having many redundant branches, which are poorly interpretable. However, this issue can be alleviated by increasing the Gini impurity (parameter \textit{min\_impurity\_decrease}) while simultaneously decreasing the maximal depth of the tree (parameter \textit{max\_depth}).

\subsection{Regression}
In contrast to predicting solvability alone, we also attempt to predict the actual clique size found by the D-Wave 2000Q device from graph and annealer features alone. Therefore, in this section, our aim is not the prediction of a binary decision (solvable, not solvable), but predicting an integer response (precisely, an integer within the range of possible clique sizes $\{0,\ldots,|V|\}$). For this task, no ground truth (via a classical solver) is needed.

As in the previous section, several regression models are suitable for this task. We chose to predict the clique size returned by D-Wave 2000Q with gradient boosting, a popular machine learning regression model. Whereas random forests build deep independent trees, gradient boosting works by constructing an ensemble of dependent shallow trees, each one improving on the previous one. Although shallow trees by themselves are usually only weak predictive models, combining (or "boosting") them in an ensemble allows for powerful machine learning models. We employed the \textit{GradientBoostingRegressor()} class \cite{gradientboosting} in \textit{scikit-learn} \cite{scikit-learn}. Similarly to the classification setting, we performed a grid search across selected parameters, and set \textit{n\_estimators=200}, and \textit{random\_state=0}. All other parameters were kept at their default values.

The graph and annealer features included in the model are the same as the ones outlined in the previous section. One caviat is worth mentioning. It is not always true that the set of vertices (given by the bitstring indicating the clique vertices with a $1$) returned by the D-Wave 2000Q annealer actually forms a clique. Therefore, to train the model, we verify first if the result returned by D-Wave 2000Q is a clique. If it is, we use the clique size in the regression, otherwise we use a value of zero.

\section{Experimental results}
\label{sec:experiments}
This section presents our results when training the machine learning models described in Section~\ref{sec:methods} on random MC instances. To make results transferable and interpretable across different MC instances being solved, we fix the embedding of a complete $65$ vertex graph onto D-Wave's Chimera architecture for all experiments.

We first consider a fixed annealing time and a fixed UTC prefactor parameter (Section~\ref{sec:fixedsetting}), and then generalize the results to random annealing times and random UTC prefactors (Section~\ref{sec:randomsetting}). For both cases, the classification and regression problems are considered.

\subsection{Fixed annealing time and fixed UTC prefactor}
\label{sec:fixedsetting}
We generate a dataset of $47000$ MC instances as described in Section~\ref{sec:methods}. In this section, the annealing time was held fixed at $100$ microseconds, and the UTC prefactor of D-Wave's \textit{uniform torque compensation} feature for computing the chain strength was empirically fixed at $0.5$, which yielded good approximate cliques with D-Wave.

\subsubsection{Classification}
\begin{figure*}
    \centering
    \includegraphics[width=\textwidth]{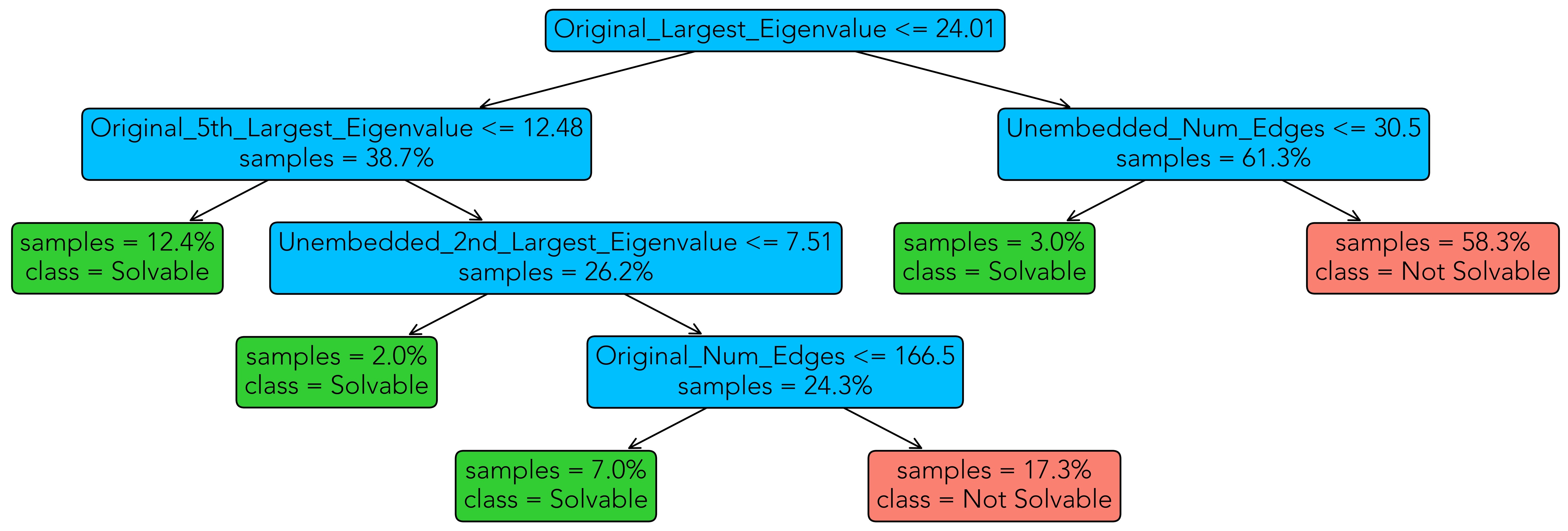}
    \caption{Decision tree for classification of MC instances into solvable and unsolvable, using the features outlined in Section~\ref{sec:methods}. Setting of fixed annealing time and fixed UTC prefactor. Left branches denote \textit{true} bifurcations, right branches denote \textit{false} bifurcations. Green leaves are solvable cases, red leaves are unsolvable cases, and inner nodes are colored in blue.\label{fig:tree_fixed}}
\end{figure*}

We start by looking at the classification problem. After fitting a decision tree classifier with \textit{scikit-learn}, we obtain the tree displayed in Figure~\ref{fig:tree_fixed}.

Several observations regarding Figure~\ref{fig:tree_fixed} are noteworthy. First, the tree is sufficiently simple to allow for a manual classification of MC instances into solvable and unsolvable ones by D-Wave 2000Q. The most decisive feature turns out to be the largest eigenvalue of the adjacency matrix of the input graph. If it is larger than the threshold given in Figure~\ref{fig:tree_fixed}, the number of edges in the unembedded problem graph seems to be a good predictor of solvability. This is sensible, since less connections in the unembedded problem tend to facilitate embedding and solving on D-Wave. However, further work is needed to understand how to the $5$'th largest eigenvalue of the adjacency matrix of the input graph comes into play. Here, a smaller $5$'th largest eigenvalue tends to facilitate solvability. Finally, the second largest eigenvalue of the unbedded problem graph (as before, smaller eigenvalues tend to facilitate solvability) and, moreover, the number of edges in the input graph (smaller instances tend to be more likely to be solvable than larger ones) tend to be good predictors of solvability---a finding which seems straightforward.

Although the importance of the $5$'th largest eigenvalue (of the adjacency matrix of the input graph) is a surprising result, the predictive power of the largest and second largest eigenvalues is sensible, since those are well known to predict a variety of structural properties of a graph, see \cite{Cvetkovic1995, lovasz}: for instance, the largest eigenvalue is related to the largest degree of a graph. In two follow-up experiments, removing only the $5$'th largest eigenvalue from the set of predictors turned out to decrease performance marginally (the accuracy decreases by 0.37\%, and the recall decreases by 0.51\%). Removing all eigenvalues apart from the largest one seems to decrease performance more substantially (the accuracy drops by 2.9\%, but the recall goes up by 0.83\%).

\begin{table}
    \centering
    \begin{tabular}{cc|cc}
        &&\multicolumn{2}{c}{Predicted}\\
        &&\multicolumn{1}{c}{Not Solvable}
        &\multicolumn{1}{c}{Solvable}\\
        \hline
        \multirow[c]{2}{*}{\rotatebox[origin=tr]{90}{Actual}}
        & Not Solvable & 3458 & 654\\[1.5ex]
        & Solvable & 97 & 497\\
        \hline
    \end{tabular}
    \caption{Confusion matrix for evaluating the decision tree (Figure~\ref{fig:tree_fixed}) on the test dataset. Setting of fixed annealing time and fixed UTC prefactor. \label{tab:conf_mat_fixed}}
\end{table}
An evaluation of the decision tree of Figure~\ref{fig:tree_fixed} on the unseen testing data yields the confusion matrix given in Table~\ref{tab:conf_mat_fixed}. We observe that our machine learning model is able to accurately classify solvable instances with a probability of around $83.7\%$, and unsolvable ones with a probability of around $84.1\%$.

\subsubsection{Regression}
Next, we consider the prediction of the clique size that the D-Wave 2000Q annealer will find on certain MC instances, as a function of the features outlined in Section~\ref{sec:methods}. All results in this section were obtained with the gradient boosting regressor of \textit{scikit-learn}.

\begin{figure*}
    \centering
    \includegraphics[width=0.44\textwidth]{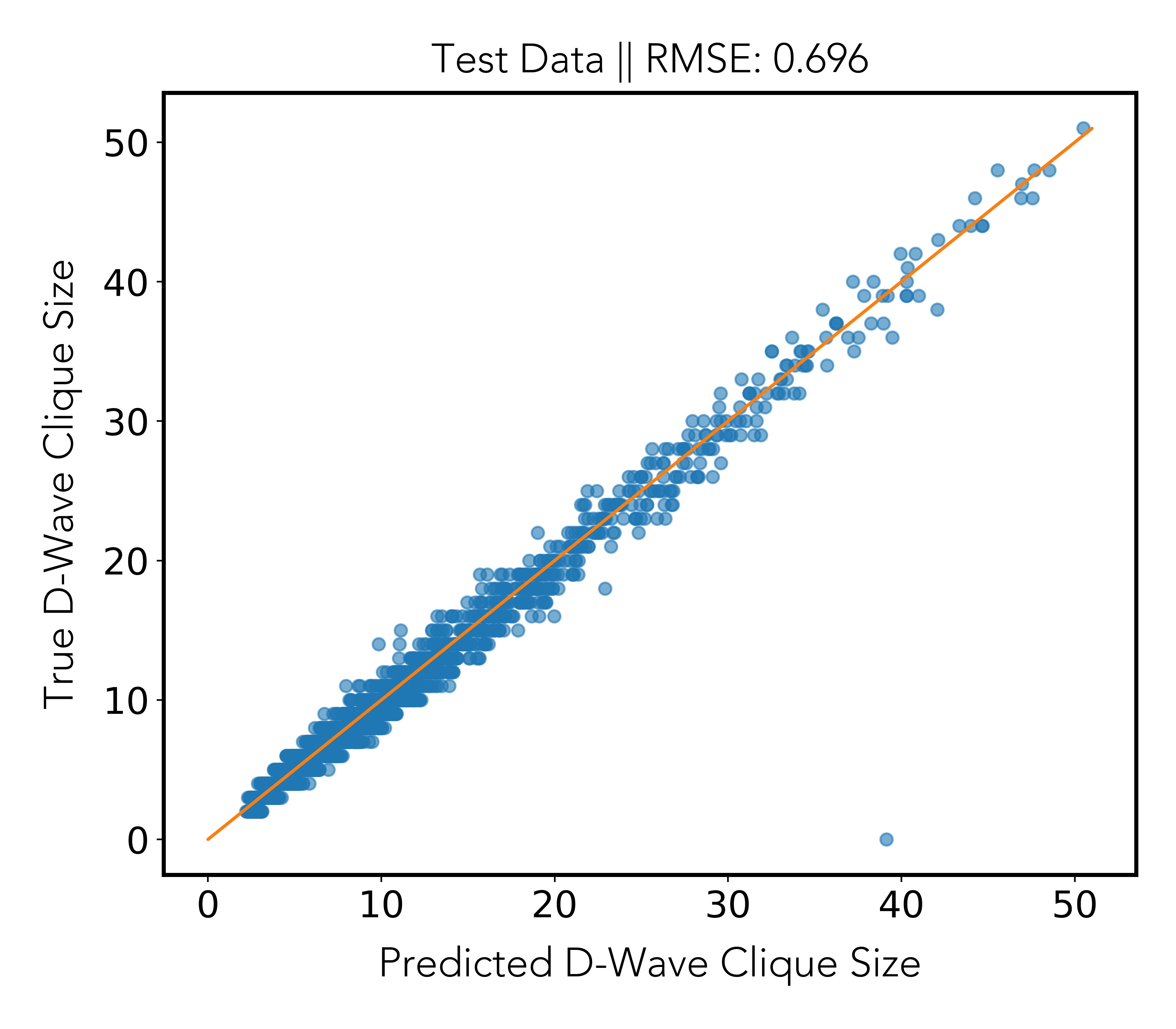}~
    \includegraphics[width=0.56\textwidth]{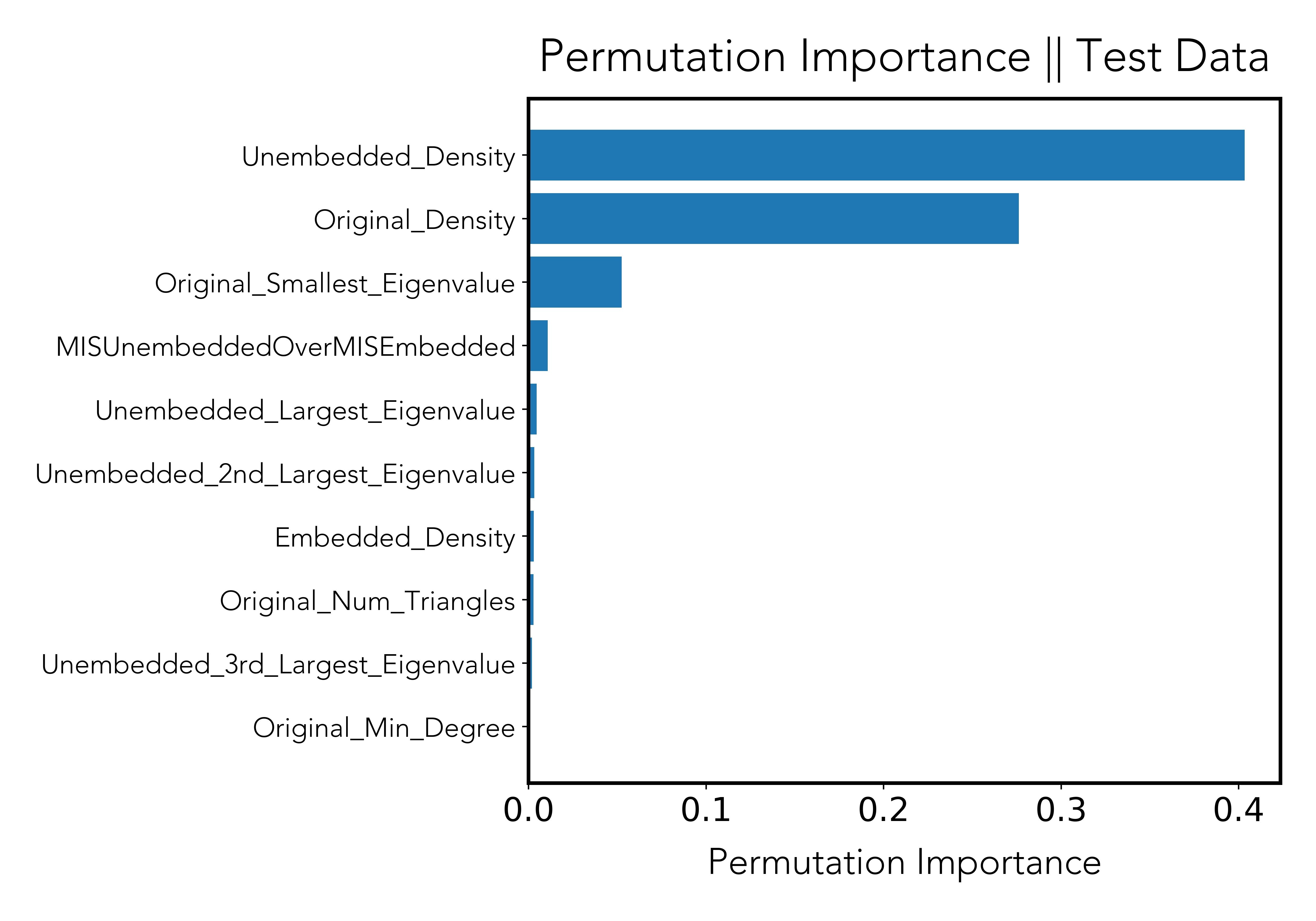}
    \caption{Regression via gradient boosting. Left: predicted D-Wave clique size vs.\ true D-Wave clique size. Right: permutation importance ranking. Setting of fixed annealing time and fixed UTC prefactor. \label{fig:regression_fixed}}
\end{figure*}
Figure~\ref{fig:regression_fixed} shows both the predicted D-Wave clique size versus the one actually found by the annealer (left plot), as well as the permutation importance ranking of the features returned by the gradient boosting algorithm (right plot). Permutation importance ranking is a means to compute the importance of each feature \cite{Breiman2001}. It works by measuring the increase in the prediction error of the model after we permuted the feature's values, which breaks the relationship between the feature and the true outcome (since breaking the relationship between a feature and the outcome makes classification easier).

We observe that, indeed, the model allows for a relatively accurate forecast of the found clique size, with a root mean square error of $0.696$ which is quite low considering the predicted values lay in the range of up to $50$. The most important features seem to be the density of the unembedded graph, the density of the original input graph, and the smallest eigenvalue of its adjacency matrix.

\subsection{Random annealing time and random UTC prefactor}
\label{sec:randomsetting}
To see if the aforementioned results generalize, we repeated the same experiment with a random annealing time (sampled uniformly at random within the interval of $[1,2000]$ microseconds), and a random UTC prefactor (sampled uniformly at random within the interval $[0.5,3]$). The dataset used for classification contained $49000$ samples and was generated as outlined in Section~\ref{sec:methods}. We again consider both classification and regression. Note that in contrast to Section~\ref{sec:fixedsetting}, where only graph features were used to train the  machine learning models, in this section we have two additional parameters in our model (annealing time and UTC prefactor), which are related to the quantum annealing algorithm.

\subsubsection{Classification}
\begin{figure*}
    \centering
    \includegraphics[width=\textwidth]{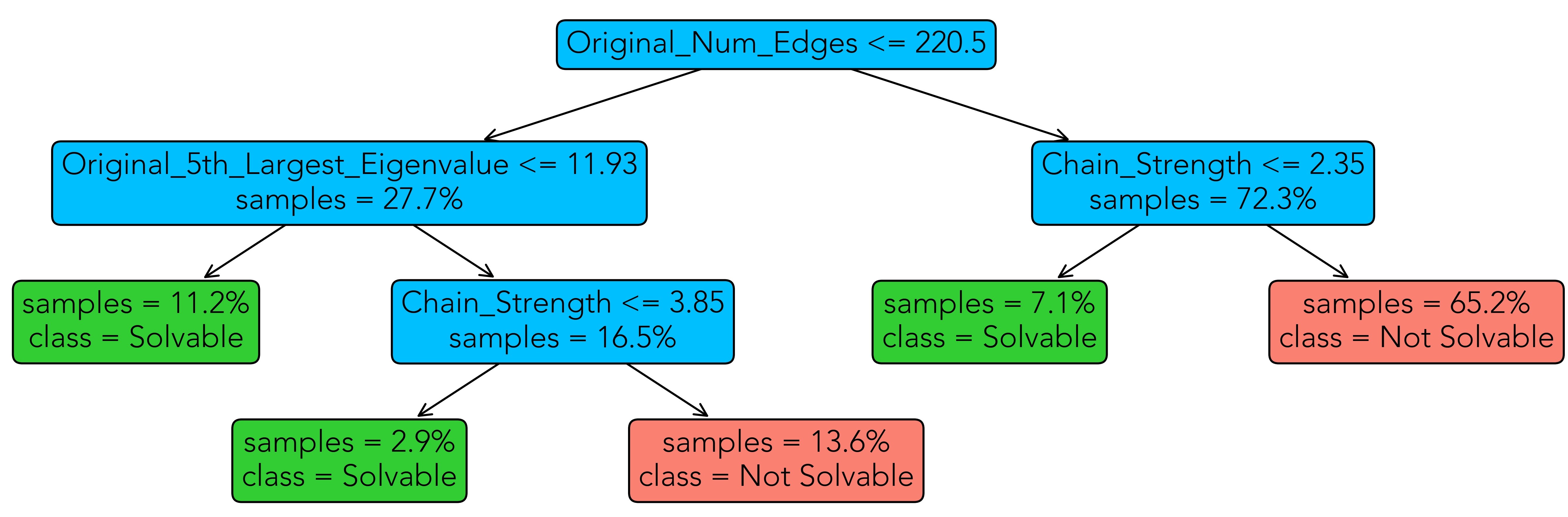}
    \caption{Decision tree for classification of MC instances into solvable and unsolvable, using the features outlined in Section~\ref{sec:methods}. Setting of random annealing time and random UTC prefactor. Left branches denote \textit{true} bifurcations, right branches denote \textit{false} bifurcations. Green leaves are solvable cases, red leaves are unsolvable cases, and inner nodes are colored in blue.\label{fig:tree_random}}
\end{figure*}
Figure \ref{fig:tree_random} shows the decision tree we obtain on the test dataset after fitting a decision tree classifier with \textit{scikit-learn}. It is similar to the one of Section~\ref{sec:fixedsetting} in that it is suitably simple to allow one to classify MC instances manually. However, the importance of the features has changed: The number of edges, though previously also included in the tree of Figure~\ref{fig:tree_fixed}, is now the first decision point. Instances with a large number of edges are likely not solvable by D-Wave, whereas most cases belonging to the left branch (number of edges at most $220$) are solvable. Similarly to Section~\ref{sec:fixedsetting}, the $5$'th largest eigenvalue of the adjacency matrix of the input graph is an important feature. New in this scenario is the importance of the chain strength for the classification, whereby small chain strengths used to embed a problem are favorable for solvability.

On the validation dataset, the decision tree of Figure~\ref{fig:tree_random} shows a similar performance to the one of the previous section. Results are summarized in Table~\ref{tab:conf_mat_random} and show that our classifier is able to classify solvable instances correctly with an accuracy of around $86.2\%$, and unsolvable ones with an accuracy of around $84.7\%$.
\begin{table}
    \centering
    \begin{tabular}{cc|cc}
        &&\multicolumn{2}{c}{Predicted}\\
        &&\multicolumn{1}{c}{Not Solvable}
        &\multicolumn{1}{c}{Solvable}\\
        \hline
        \multirow[c]{2}{*}{\rotatebox[origin=tr]{90}{Actual}}
        & Not Solvable & 3731 & 672\\[1.5ex]
        & Solvable & 68 & 425\\
        \hline
    \end{tabular}
    \caption{Confusion matrix for evaluating the decision tree (Figure~\ref{fig:tree_random}) on the test dataset. Setting of random annealing time and random UTC prefactor.\label{tab:conf_mat_random}}
\end{table}

\subsubsection{Regression}
\begin{figure*}
    \centering
    \includegraphics[width=0.44\textwidth]{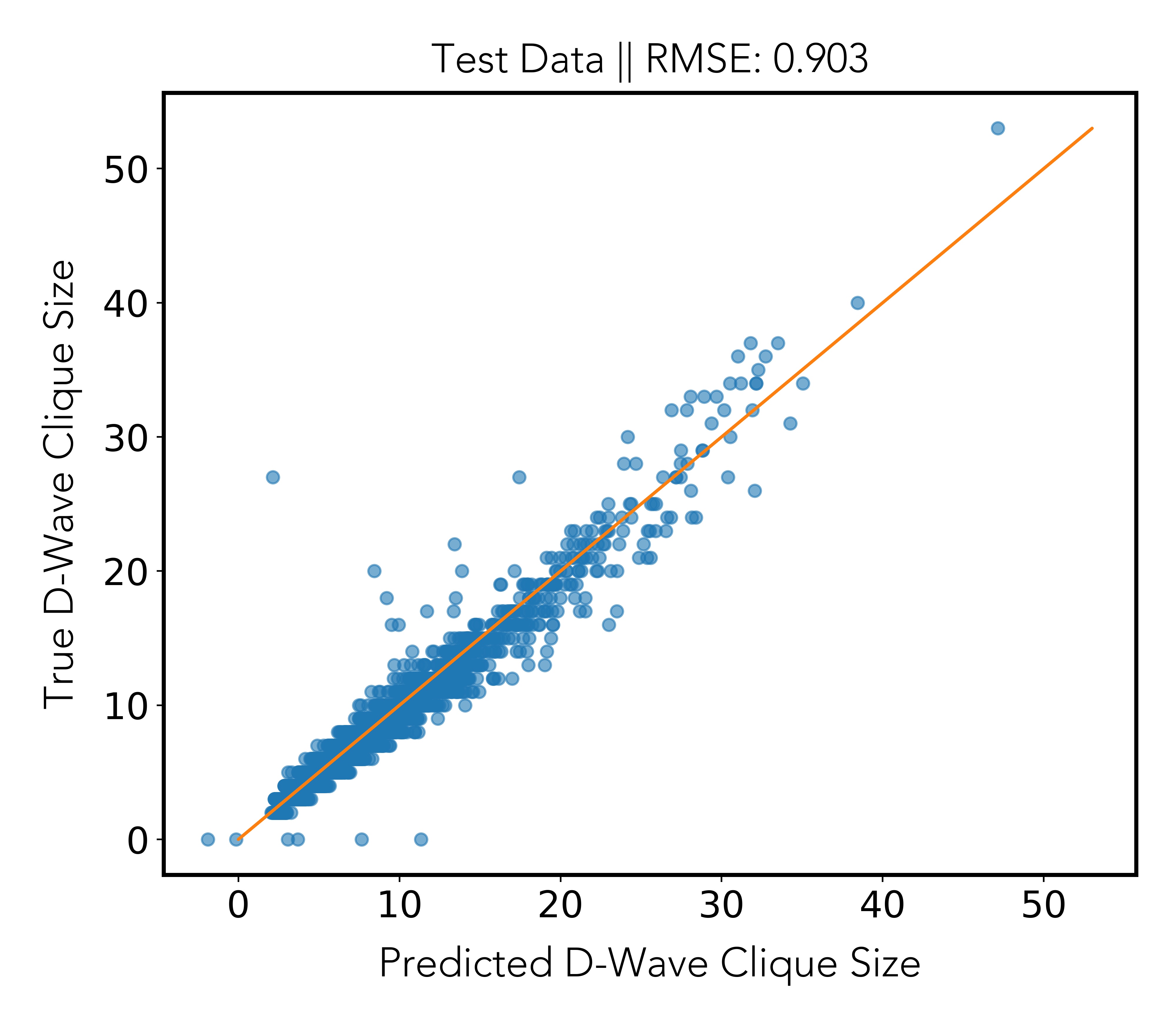}~
    \includegraphics[width=0.56\textwidth]{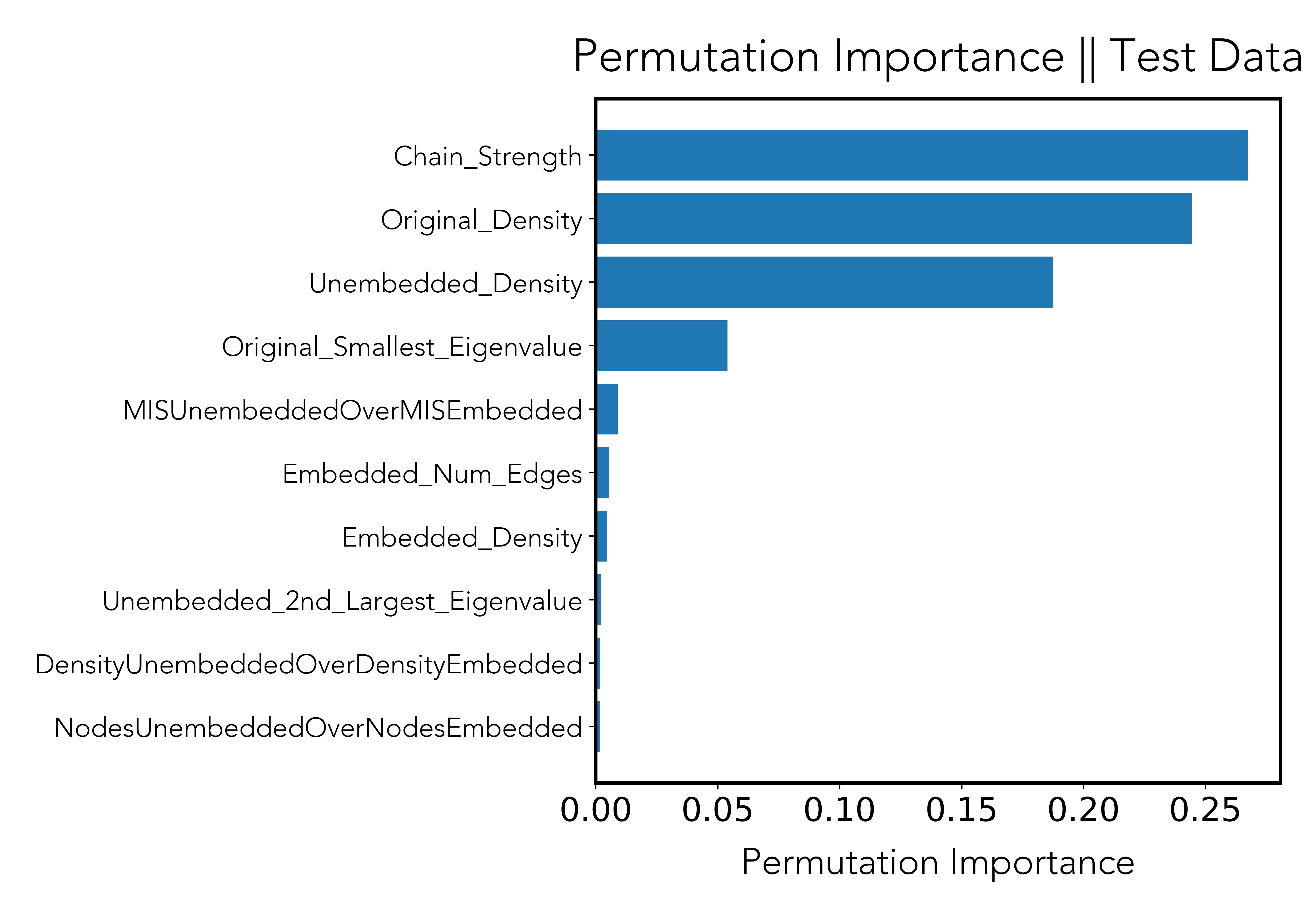}
    \caption{Regression via gradient boosting. Left: predicted D-Wave clique size vs.\ true D-Wave clique size. Right: permutation importance ranking. Setting of random annealing time and random UTC prefactor.\label{fig:regression_random}}
\end{figure*}

Finally, we repeat the machine learning regression with the help of the gradient boosting regressor of \textit{scikit-learn}. Figure~\ref{fig:regression_random} shows the results of this experiment.

On the test dataset, we are able to very accurately predict the clique size which D-Wave 2000Q will find, achieving a low root mean square error of $0.903$ (Figure~\ref{fig:regression_random}, left). The most important features for regression seem to be chain strength, the density of the original input graph, and the density of the unembedded graph. The smallest eigenvalue of the adjacency matrix of the original input graph seems to play a minor role.

\section{Discussion}
\label{sec:discussion}
In this contribution, we aim to understand some of the factors contributing to the hardness of a problem instance sent to the D-Wave 2000Q annealer. We focus on the MC problem, and train several machine learning models on several thousand randomly generated input problems with the aim to learn features to (a) predict if D-Wave 2000Q will be able to solve an instance of MC to optimality, and (b) predict the size of the clique which the D-Wave 2000Q device will find.

We show that indeed, for specific problem of predicting the clique size, a relatively small number of features of the input problem, its QUBO formulation and its embedding onto the D-Wave Chimera architecture seems to suffice to predict solubility with high precision. Those features can be summarized in a simple decision tree, which even allows one to manually classify MC instances. A regression analysis demonstrated that the clique size the D-Wave 2000Q will find can likewise be predicted with a low root mean square error.

This article leaves scope for a variety of avenues for future work. For instance, it is unknown how these results generalize to other NP-hard problems, how well prediction will work on future D-Wave machines (or even other quantum annealers), and it remains to be investigated if the presented results can be improved by fitting more sophisticated machine learning models.

\section*{Acknowledgments}
This work has been supported by the US Department of Energy through the Los Alamos National Laboratory. Los Alamos National Laboratory is operated by Triad National Security, LLC, for the National Nuclear Security Administration of U.S. Department of Energy (Contract No.~89233218CNA000001) and by the Laboratory Directed Research and Development program of Los Alamos National Laboratory under project numbers 20190065DR and 20180267ER.


\end{document}